\documentclass[11pt]{article}
\usepackage{amsmath,amssymb,color,epsf}

\makeatletter
\@addtoreset{equation}{section}
\makeatother

\newcommand{\hoch}[1]{$\, ^{#1}$}

\newcommand{\be}{\begin{equation}}
\newcommand{\ee}{\end{equation}}
\newcommand{\bea}{\setlength\arraycolsep{2pt} \begin{eqnarray}}
\newcommand{\eea}{\end{eqnarray}}
\newcommand{\nn}{\nonumber}

\def\ft#1#2{{\textstyle{\frac{\scriptstyle #1}{\scriptstyle #2} } }}
\def\fft#1#2{{\frac{#1}{#2}}}

\def\0{{\sst{(0)}}}
\def\1{{\sst{(1)}}}
\def\2{{\sst{(2)}}}
\def\3{{\sst{(3)}}}
\def\4{{\sst{(4)}}}
\def\5{{\sst{(5)}}}
\def\6{{\sst{(6)}}}
\def\7{{\sst{(7)}}}
\def\8{{\sst{(8)}}}
\def\sst#1{{\scriptscriptstyle #1}}
\def\oneone{\rlap 1\mkern4mu{\rm l}}

\def\del{{\partial}}

\def\cG{{{\cal G}}}
\def\cA{{{\cal A}}}
\def\cF{{{\cal F}}}
\def\cJ{{{\cal J}}}
\def\cQ{{{\cal Q}}}
\def\cP{{{\cal P}}}
\def\cB{{{\cal B}}}
\def\cG{{{\cal G}}}

\begin{document}

%%%%%%%%%%%%%%%%%%%%%%%%%%%%%%%%%%%%%%%%

\begin{flushright}
\hfill{ \
MIFPA-13-28\ \ \ \ }
 %\hfill{
%\bf hep-th/yymmnnn}
\end{flushright}

\vspace{25pt}
\begin{center}
{\Large {\bf Thermodynamics of Magnetised Kerr-Newman Black Holes}
}

\vspace{30pt}

{\Large
G.W. Gibbons\hoch{1}, Yi Pang\hoch{2} and C.N. Pope\hoch{1,2}
}

\vspace{10pt}

\hoch{3}{\it DAMTP, Centre for Mathematical Sciences,
 Cambridge University,\\  Wilberforce Road, Cambridge CB3 OWA, UK}

\vspace{10pt}

\hoch{2} {\it George P. \& Cynthia Woods Mitchell  Institute
for Fundamental Physics and Astronomy,\\
Texas A\&M University, College Station, TX 77843, USA}

\vspace{20pt}

\underline{ABSTRACT}
\end{center}
\vspace{15pt}

The thermodynamics of a  magnetised Kerr-Newman black hole is
studied to all orders in the appended magnetic field $B$.  The
asymptotic properties of the metric and other fields are dominated by the
magnetic flux that extends to infinity along the axis, leading to subtleties
in the calculation of conserved quantities such as the angular momentum
and the mass.  We present a detailed discussion of the implementation
of a Wald-type procedure to calculate the angular momentum, showing how
ambiguities that are absent in the usual asymptotically-flat case
may be resolved by the requirement of gauge invariance.
We also present a formalism from which we are able to 
obtain an expression for the mass of the magnetised black holes.
The expressions for the mass and the angular momentum are shown to be
compatible with the first law of thermodynamics and a
Smarr type relation.  Allowing the appended magnetic field $B$ to vary
results in an extra term in the first law of the form $-\mu dB$ where
$\mu$ is interpreted as an induced magnetic moment. Minimising the
total energy with respect to the total charge $Q$
at fixed values of the angular momentum and energy of the
seed metric allows an investigation of Wald's process.
The Meissner effect is shown to hold for electrically
neutral extreme black holes.  We also present a derivation of the
angular momentum for black holes in the four-dimensional STU model,
which is ${\cal N}=2$ supergravity coupled to three vector multiplets.

\thispagestyle{empty}

\pagebreak
\voffset=-40pt
\setcounter{page}{1}

\tableofcontents

\addtocontents{toc}{\protect\setcounter{tocdepth}{2}}

\newpage

\section{Introduction}

Understanding the behaviour of rotating black holes
and their ergoregions when  immersed in
magnetic fields forms  a central part of
astrophysical theories of quasars, active galactic nuclei
and other objects containing black holes (see, for example,
\cite{KKL,BlandfordZnajek,Bicak1}).
Typically the magnetic fields, which contribute
a negligible amount of energy density near the black hole, are treated
as test fields on the background of the  Kerr  solution
of the vacuum Einstein  equations, as, for example, in the work of Wald 
\cite{Wald0}. While perfectly justified
astrophysically, it is not without interest  to treat the energy
exchange between black hole and magnetic field at the fully non-linear
level, and in particular to ask to what extent  the ideas of
black hole thermodynamics, which have proved so useful in the study
of  quantum processes near black holes, may be extended to
this more general setting. This seems especially appropriate
since rotating black holes are believed to  drag magnetic field lines, inducing
electric fields to flow and hence currents to flow.
For sufficiently strong magnetic fields  this may lead to the breakdown
of the vacuum due to pair creation \cite{Gibbons,Putten1,Putten2}.

These thoughts motivated a recent  study  \cite{gimupo} of the exact metric
and  electromagnetic field  of a magnetized Kerr-Newman black hole,
constructed using solution-generating methods pioneered by Ernst \cite{Ernst}.
Contrary to the widespread  belief that
the asymptotic metric was approximately static and Melvin-like,
it was found that  generically  the metric has ergoregions that extend
all the way to infinity. The complicated  nature of the metric
at infinity presented difficulties in evaluating the
total energy and angular momentum of the system and in the treatment
of the thermodynamics.  In this paper, we are able partially to
overcome these problems and to present expressions
for the total angular momentum $J$ and total energy $E$ of the system,
together with  a form of the relevant Smarr relation and
and first law in which variations of the appended magnetic
field $B$ are fully taken into account.  (The thermodynamics of the
Schwarzschild-Melvin black hole were discussed in \cite{radu}.)

The plan of the paper is as follows. In section 2 we discuss the Wald
procedure for evaluating the total angular momentum $J$, and some
subtleties that can arise in cases such as the magnetised black holes
that were not present for the asymptotically-flat geometries
considered by Wald. These lead to potential ambiguities in the definition
of the angular momentum.  We argue that these may be resolved by a careful 
consideration of the behaviour of the conserved angular momentum under
gauge transformations.  In section 3 we show how the electric charge
and the angular momentum may
be conveniently evaluated by first performing a Kaluza-Klein reduction on
the azimuthal coordinate $\phi$, and then expressing the conserved Wald
charge in terms of three-dimensional quantities.

The definition of the mass of a black hole in an external magnetic
field is also somewhat problematical, on account of the unusual asymptotic
behaviour of the metric and the other fields. We discuss this in section 4,
where we present a formalism, again based on the Kaluza-Klein reduction
to three dimensions, within which we are able to obtain an expression
for the mass.
In section 5 we evaluate our general expressions for the angular momentum and
the mass in the case of the magnetised Kerr-Newman black holes.  We show
that these results are consistent with the first law of thermodynamics, in
the case that the appended magnetic field $B$ is held fixed.  In doing so,
we essentially use the first law to derive expressions for the angular
velocity $\Omega$ and the electrostatic potential $\Phi$.\footnote{To 
be precise, $\Omega$ and $\Phi$ represent the {\it differences}
$\Omega=\Omega_H-\Omega_\infty$ and $\Phi=\Phi_H-\Phi_\infty$ between
the values on the horizon and the values at infinity. $\Omega_H$ and
$\Phi_H$ are easily computed directly, but the asymptotics of the
magnetised black hole solutions make it difficult to define
$\Omega_\infty$ and $\Phi_\infty$ directly.}
We then
extend the discussion, treating $B$ also as a thermodynamic variable
by introducing an extra contribution $-\mu dB$ in the first law, where
$\mu$ has an interpretation as an induced magnetic moment.
The explicit expressions for $\Omega$, $\Phi$ and $\mu$
may be calculated exactly, but their forms are rather
complicated. However they simplify considerably if one works
to low orders in $B$ or $q$.     We show also that the thermodynamic
variables in the extended system obey a Smarr-type relation.

In section 6, we attempt to compare our thermodynamic
formalism with some work  of Wald \cite{Wald,Wald2}.
We  minimize the total energy $E$
with respect to the total charge $Q$,
at fixed values of the  energy,  angular momentum
and magnetic field of the black hole.
Our result resembles that of Wald in general form, but differs in detail.

In section 7 we examine some further properties of the
magnetised black holes, including  the  Meissner effect 
whereby as one approaches extremality, the magnetic  flux
penetrating the horizon vanishes. In other words, flux is 
expelled \cite{Bicak}.
We find, that if the total charge on the hole $Q$ vanishes,
then the magnetic field on the horizon does indeed vanish as one
approaches extremality, consistent with earlier work.

In section 8 we extend our discussion of the conserved angular momentum to
the case of the STU supergravity model, which comprises four-dimensional
${\cal N}=2$ supergravity coupled to three vector multiplets.  We
apply our results to the case of the magnetisation of certain
4-charge static black holes that have been investigated recently in
\cite{cvgiposa}.  The paper ends with conclusions in section 9.

\section{Conserved Charges in Einstein-Maxwell Theory}

    Here we present a discussion of some aspects of the Wald procedure
\cite {Wald2} for calculating conserved charges, applied to the case of
the four-dimensional Einstein-Maxwell theory.  Our motivation for
doing so will be as part of an investigation of the thermodynamics of
the magnetised
Kerr-Newman black holes that were recently studied in \cite{gimupo}.
We shall find that some subtleties arise in this context that make
it necessary to pay close attention to some of the details of the
Wald procedure.

Starting from the Einstein-Maxwell Lagrangian
%%%%%
\be
{\cal L}_4=\fft{1}{16 \pi}\, (R\, {*\oneone}- 2{*F}\wedge F)
\,,\label{originallag}
\ee
%%%%%
and following a calculation developed by Wald \cite{Wald,Wald2}, one can use
the Noether procedure to derive a current ${\cal J}$, given
by
%%%%%
\be
{\cal J} = -d{*d}\xi - 4 {*F}\wedge d(\xi^\mu\, A_\mu)\,,\label{cJexp}
\ee
%%%%%
where $\xi=\xi_\mu\, dx^\mu$ and $\xi^\mu\, \del_\mu$ is a Killing
vector.\footnote{We shall present a detailed derivation in section
\ref{stusec} of the analogous result in the more complicated context of
the STU supergravity model.}
Since $d\cJ=0$, we can write $\cJ=-d\cP$ and hence derive the conserved charge
%%%%%
\be
\cQ[\xi]=\frac{1}{16\pi G}\int_{S^2} \cP  \,.\label{Qcon1}
\ee
%%%%%
From this point on we shall work in units where $G=1$.

  One way to obtain a local expression for $\cP$ is to note that the Maxwell
equation $d{*F}=0$ allows us to extract an exterior derivative from
the second term in (\ref{cJexp}) and write
%%%%%
\be
\cP = {*d}\xi +4 {*F}\, (\xi^\mu\, A_\mu)\,.\label{waldP}
\ee
%%%%%
This is the the form in which the conserved charge was obtained in
\cite{Wald}.

  An objection one may raise to the expression (\ref{waldP}) is that it is
not invariant under gauge transformations of $A$.  Specifically, if
we send $A\longrightarrow A + d\lambda$, then we shall have
%%%%%
\be
\cP\longrightarrow \cP + 4 {*F}\, (\xi^\mu\, \del_\mu\lambda)\,.
\ee
%%%%%
Since the Killing vector $\xi^\mu$ generates a symmetry of the solution,
it follows that the Lie derivative of $F$ will vanish, ${\mathfrak L}_\xi F=0$.
We may assume that a gauge choice for $A$ is made so
that ${\mathfrak L}_\xi A=0$ also.  However,
there can still remain a residual gauge freedom that preserves
this choice, namely when the gauge parameter $\lambda$ satisfies
%%%%%
\be
\xi^\mu\, \del_\mu\lambda = c\,,\label{residuals}
\ee
%%%%%
where $c$ is a constant.  This can be seen from the fact that
gauge transformations preserving ${\mathfrak L}_\xi\, A=0$ 
must satisfy ${\mathfrak L}_\xi\,d\lambda
=(i_\xi\, d + d\, i_\xi)d\lambda = d\, i_\xi\, d\lambda =
d(\xi^\mu\, \del_\mu\lambda)=0$.\footnote{Here $i_\xi$ denotes the
interior product of $\xi=\xi^\mu\del_\mu$ with a $p$-form
$\omega=(1/p!)\, \omega_{\mu_1\cdots\mu_p}\, dx^{\mu_1}\wedge\cdots\wedge
dx^{\mu_p}$.
Its action is defined by $i_\xi\omega= (1/(p-1)!)\, \xi^{\mu_1}\,
\omega_{\mu_1\cdots\mu_p}\, dx^{\mu_2}\wedge\cdots\wedge
dx^{\mu_p}$. Note that if $\omega$ and $\nu$ are a $p$-form and a $q$-form,
then $i_\xi(\omega\wedge\nu)= (i_\xi \omega)\wedge \nu +
  (-1)^p\, \omega\wedge (i_\xi \nu)$.  The Lie derivative of any $p$-form
is given by ${\mathfrak L}_\xi\, \omega= 
(di_\xi + i_\xi d)\omega$.\label{idef}}
The conserved charge in (\ref{Qcon1})
will then undergo a gauge transformation of the form
%%%%%
\be
\cQ[\xi] \longrightarrow \cQ[\xi] +  c\, Q\,,
\ee
%%%%%
where $Q=1/(4\pi)\, \int{*F}$ is the electric charge.

   An alternative way of extracting an exterior derivative from
the expression (\ref{cJexp}) for $\cJ$ is to introduce a dual gauge
potential $\widetilde A$ such that ${*F}\equiv\widetilde F = d\widetilde A$,
and then write $\cJ=-d\widetilde\cP$, where
%%%%%
\be
\widetilde\cP = {*d}\xi +4 \widetilde A\wedge d(\xi^\mu\, A_\mu)\,.
\label{gaugeinvP}
\ee
%%%%%
It is evident that the corresponding conserved charge $\widetilde\cQ[\xi]$
obtained by substituting $\widetilde\cP$ into (\ref{Qcon1}) will
be invariant under the residual gauge transformations of $A$, which
satisfy (\ref{residuals}).

  Our principle interest in this paper will be to apply the Wald construction
to the calculation of the angular momentum.
  The two alternative expressions (\ref{waldP}) and (\ref{gaugeinvP}) for
a 2-form whose exterior derivative gives ${\cal J}$ then essentially
correspond to the standard Wald expressions that one obtains by using
either the original Lagrangian (\ref{originallag}) (leading to (\ref{waldP}))
or else the dual Lagrangian\footnote{The dual Lagrangian can be
obtained by adding a Lagrange multiplier term to (\ref{originallag}) and
writing
%%%%%
\be
{\cal L}= \fft1{16\pi G}\, (R\, {*\oneone} -2{*F}\wedge F + 4d\widetilde A
\wedge F)\,,
\ee
%%%%%
where now $F$ and $\widetilde A$ are viewed as fundamental fields.
The equation of motion for $\widetilde A$ implies the usual Bianchi
identity for $F$.  If instead we eliminate $F$ via its (algebraic)
equation of motion, we obtain (\ref{duallag}).  The original and the
dual Lagrangian differ on-shell by the total derivative term
$-4(d\widetilde A\wedge dA)/(16\pi G)$.}
%%%%%
\be
\widetilde{\cal L}_4 = \fft{1}{16\pi G}\, (R\, {*\oneone} -
   2{*\widetilde F}\wedge\widetilde F)\,,\label{duallag}
\ee
%%%%%
where $\widetilde F= {*F}=d\widetilde A$.  To see this, consider the
analogue of the Wald expression (\ref{waldP}) that one would derive
from the dual Lagrangian (\ref{duallag}):
%%%%%
\be
\cP_{\rm dual} = {*d}\xi +4 {*\widetilde F}\, (\xi^\mu\, \widetilde A_\mu)
\,.\label{waldPdual}
\ee
%%%%%
The difference between this and $\widetilde \cP$ (defined in (\ref{gaugeinvP}))
is therefore
%%%%%
\be
\cP_{\rm dual}-\widetilde \cP= 4{*\widetilde F}\,\, i_\xi \widetilde A
  - 4\widetilde A\wedge di_\xi A\,,
\ee
%%%%%
Assuming that $\xi^\mu$ is a
Killing vector, so that the Lie derivative of the field strength $F$
in a solution vanishes, ${\mathfrak L}_\xi F =(di_\xi + i_\xi d)F=0$,
we may choose gauges where ${\mathfrak L}_\xi A=0$ and 
${\mathfrak L}_\xi\widetilde A=0$.  In
particular, this means that $di_\xi A=-i_\xi d A = -i_\xi F$.  Using also
that ${*\widetilde F}=-F$, we see that
%%%%%
\bea
\cP_{\rm dual}-\widetilde \cP&=& -4 i_\xi \widetilde A\, F + 4\widetilde A
  \wedge i_\xi F\,,\nn\\
&=& -4i_\xi(\widetilde A\wedge F)\,.\label{Pdiff}
\eea
%%%%%
In particular, if we consider the case when $\xi=\del/\del\phi$
is the
Killing vector that generates azimuthal rotations, then it follows from
the final line of (\ref{Pdiff}) that $(\cP_{\rm dual}-\widetilde \cP)$
has no pullback onto the 2-sphere over which we integrate to obtain
a conserved charge. This means that
$\cP_{\rm dual}$ and $\widetilde \cP$ would give
identical expressions for the angular momentum.

  In the following section,
we shall discuss the dimensional reduction of the theory, and
its solutions, on the azimuthal Killing vector $\del/\del\phi$.  This
will provide us with a formalism that is particularly well adapted to
computing the angular momentum for the solutions we are interested in.

\section{Conserved Charge, Angular Momentum and Mass via Dimensional Reduction}

   A convenient way of calculating the conserved charges is to perform
a Kaluza-Klein dimensional reduction on the $\phi$ coordinate.  Thus we
write\footnote{Note that the reduction ansatz for $A$ is compatible
with the partial gauge condition ${\mathfrak L}_\xi A=0$ 
that we discussed previously,
since $i_\xi A= \chi$ and $i_\xi dA= -d\chi$, so $(di_\xi + i_\xi d)A=0$.}
%%%%%
\be
d{s}^2_4 = e^{2\varphi}d\bar s^2_3+e^{-2\varphi}(d\phi+2\bar{\cal A})^2\,,
\qquad {A}=\bar A+ \chi(d\phi+2\bar {\cal A})\,,\label{kkred}
\ee
%%%%%
where, whenever there is an ambiguity, we place a ``bar'' on three-dimensional
quantities to distinguish them from the unbarred four-dimensional ones.
Note that $F=\bar F + d\chi\wedge (d\phi+2\bar\cA)$.
The equations of motion for the three-dimensional fields then follow from
the dimensionally-reduced Lagrangian
%%%%%
\bea
{\cal L}_3&=&\fft{\Delta\phi}{16\pi}\, \sqrt{-\bar g} \,\Big[\bar R-
 2(\partial\varphi)^2-2e^{2\varphi}(\partial\chi)^2-
  e^{-4\varphi}\bar{\cal F}^2-e^{-2\varphi} \bar F^2\Big]\,, \\
\bar {\cal F}&=&d\bar{\cal A},
 \quad \bar F=d\bar A+2\chi d\bar{\cal A}\,,
\eea
%%%%%
where $\Delta\phi$ is the period of the azimuthal coordinate $\phi$.
The equations of motion for $\bar A$ and $\bar{\cal A}$ imply that
we can write
%%%%%
\be
e^{-2\varphi}\, {\bar*\bar F}=d\psi,\qquad e^{-4\varphi}\, {\bar*\bar {\cal F}}
   =d\sigma-2\chi d\psi\,.\label{Fduals}
\ee
%%%%%
Here, $\psi$ and $\sigma$ are the axionic scalar duals of the 1-form
potentials $\bar A$ and $\bar{\cal A}$.

\subsection{Conserved electric charge}

  Since ${*F}=e^{-2\varphi}\, {\bar * \bar F}\wedge (d\phi +2\bar\cA) +
  e^{2\varphi}\, {\bar *d\chi}$, the  conserved electric charge is given by
%%%%%
\bea
 Q &=& \fft1{4\pi} \int_{S^2} {*F} = \fft{\Delta\phi}{4\pi}\,
  \int e^{-2\varphi} \, {\bar *\bar F}\,,\nn\\
&=& \fft{\Delta\phi}{4\pi} \, \int d\psi = \fft{\Delta\phi}{4\pi}\,
  \Big[\psi\Big]_{\theta=0}^{\theta=\pi}\,.\label{charge}
\eea
%%%%%
Note that here, and henceforth, we are allowing for the possibility
that the period $\Delta\phi$ of the azimuthal coordinate $\phi$ might
be different from $2\pi$.  In particular, this happens in the case of
the magnetised black hole solutions that we shall be considering in
this paper.

\subsection{Conserved angular momentum}

  We first calculate the angular momentum $J=\cQ[\xi]$ using (\ref{Qcon1})
with $\cP$ given by (\ref{waldP}), and with
$\xi=\del/\del\tilde\phi$ where $\tilde\phi$ is the
canonically-normalised azimuthal angular coordinate with period $2\pi$.
It will, in general, be related to $\phi$ by $\phi=\alpha\tilde\phi$, with
$\alpha=\Delta\phi/(2\pi)$.  As a 1-form, $\xi$ will be given, in terms of the
three-dimensional quantities, by
%%%%%
\be
\xi = \alpha\, e^{-2\varphi}\, (d\phi+2\bar\cA)\,,
\ee
%%%%%
and furthermore $\xi^\mu A_\mu= \alpha \chi$, so from (\ref{waldP})
%%%%%
\be
\cP = *\Big[2\alpha e^{-2\varphi}\, \bar\cF + 4\alpha \chi \bar F
   -2\alpha(e^{-2\varphi}\, d\varphi -2\chi d\chi)\wedge (d\phi+2\bar\cA)\Big]
\,.
\ee
%%%%%
Thus using (\ref{Fduals}) we have
%%%%%
\bea
\cP &=& 2\alpha (e^{-4\varphi}\, {\bar *\bar \cF} + 2\chi\, e^{-2\varphi}\,
  {\bar *\bar F})\wedge (d\phi+2\bar\cA) -2\alpha({\bar *d}\varphi
    -2\chi\, e^{2\varphi}\, {\bar *d}\chi)\,,\nn\\
&=& 2\alpha d\sigma\wedge(d\phi+2\bar\cA) -2\alpha({\bar *d}\varphi
    -2\chi\, e^{2\varphi}\, {\bar *d}\chi) \,.
\eea
%%%%%
Only the first term has a non-zero pullback onto the 2-sphere, and so 
this gives a conserved angular momentum
%%%%%
\be
J = \fft1{16\pi} \int_{S^2} {\cP} = \fft{\alpha\Delta\phi}{8\pi} \int d\sigma=
     \fft{(\Delta\phi)^2}{16\pi^2}\, \Big[\sigma\Big]_{\theta=0}^{\theta=\pi}
\,.\label{angmom}
\ee
%%%%%

   As we discussed in section 2, a different choice for the definition of
the angular momentum is
to perform a dualisation of the four-dimensional field strength $F$,
and work instead with $\widetilde F= {*F}$ as the fundamental
electromagnetic field strength. As discussed in \cite{gimupo}, in the
three-dimensional language this dualisation amounts to interchanging the
three-dimensional fields $\chi$ and $\psi$.  At the same time, the
field $\sigma$ must be redefined, so that in the dual formulation we shall have
tilded fields given in terms of the original ones by \cite{gimupo}
%%%%%
\be
\widetilde\chi=\psi\,,\qquad \widetilde\psi=\chi\,,\qquad
\tilde\sigma = \sigma -2\chi\, \psi\,.\label{duality}
\ee
%%%%%
It follows that in this dualised formalism, the angular momentum defined
in (\ref{angmom}) would be replaced by
%%%%%
\be
\widetilde J =
\fft{(\Delta\phi)^2}{16\pi^2}\, \Big[\tilde\sigma\Big]_{\theta=0}^{\theta=\pi}
\,.\label{tangmom}
\ee
%%%%%

It is instructive to look at the behaviour of the
two expressions under gauge transformations.
   The quantity ${\cal P}$ defined in (\ref{Qcon1}) which we used in order to
calculate the angular momentum (\ref{angmom})
is in general gauge dependent, since the
potential $A_\mu$ appears explicitly in its construction.  This can
be seen in the three-dimensional language as follows.  If we perform a
gauge transformation $A\longrightarrow A'=A+d\lambda$ on the four-dimensional
gauge potential, then this will be compatible with the Kaluza-Klein reduction
ansatz (\ref{kkred}) for $A$ provided that $\lambda$ is restricted to
have the form
%%%%%
\be
\lambda = \bar\lambda + c\, \phi\,,\label{gaugetrans}
\ee
%%%%%
where $\bar\lambda$ depends only on the three-dimensional coordinates
and $c$ is a constant.  Specifically, comparing with the reduction
ansatz
%%%%%
\be
A' =\bar A' + \chi'\, (d\phi + 2\bar\cA')\,,
\ee
%%%%%
we see that the three-dimensional fields will transform as
%%%%%
\be
\chi'= \chi+c\,,\qquad \bar A'= \bar A-2 c \,\bar\cA + d\bar\lambda\,,\qquad
\bar \cA' =\bar\cA\,.\label{chiAtrans}
\ee
%%%%%
Since $\bar F= d\bar A + 2\chi\, d\bar\cA$, it follows that
%%%%%
\be
\bar F' = d\bar A' + 2\chi'\, d\bar\cA = d\bar A+ 2\chi\, d\bar\cA=\bar F\,,
\ee
%%%%%
and therefore from (\ref{Fduals}) we see that
%%%%%
\be
\psi'=\psi\,.\label{psitrans}
\ee
%%%%%
Since we also have $\bar\cF'=\bar\cF$ it also follows from (\ref{Fduals})
that
%%%%%
\be
d\sigma'- 2\chi'\, d\psi' = d\sigma - 2\chi\, d\psi\,,
\ee
%%%%
and so using $\chi'= \chi+c$ and $\psi'=\psi$ we see that
%%%%%
\be
\sigma'= \sigma + 2c\, \psi\,.\label{sigtrans}
\ee
%%%%%

   It follows from (\ref{sigtrans})  that if we perform the gauge
transformation in (\ref{gaugetrans}) that is parameterised by the
constant $c$, then the angular momentum given by (\ref{angmom}) will
transform to
%%%%%
\be
J' = J + c\, Q\, \fft{\Delta\phi}{2\pi}\,,\label{Jtrans1}
\ee
%%%%%
where $Q$ is the conserved electric charge given by (\ref{charge}).

   If, on the other hand, we consider the angular momentum $\widetilde J$
defined by (\ref{tangmom}), then we see that under the gauge transformations
parameterised by $c$ we shall have
%%%%%
\be
\tilde\sigma' = \sigma' - 2\chi'\, \psi'= \sigma + 2 c\,\psi -
2\chi\, \psi -2c\, \psi= \sigma-2\chi\, \psi=\tilde\sigma,
\ee
%%%%%
and so $\widetilde J$ is gauge invariant.  To be more precise, the
expression (\ref{tangmom}) for $\widetilde J$ is invariant under gauge
transformations of the original potential $A$.  Conversely, the
expression (\ref{angmom}) for $J$ is invariant under gauge transformations
of the {\it dual} potential $\widetilde A$.   Correspondingly, $J$ does
depend on gauge transformations of $A$, whilst $\widetilde J$ depends
on gauge transformations of $\widetilde A$.

  In the reduced three-dimensional description, the residual gauge
transformations of the dual potential $\widetilde A$, preserving
${\mathfrak L}_\xi\, \widetilde A=0$, correspond to sending
%%%%%
\be
\psi\longrightarrow \psi + b\,,\qquad 
\chi\longrightarrow \chi\,,\qquad
\sigma\longrightarrow \sigma\,,\label{dualtrans}
\ee
%%%%%
where $b$ is a constant parameter. This implies that the
$\tilde\sigma\equiv \sigma-2 \chi\, \psi$, which is invariant under 
the original residual gauge transformations, will transform as
%%%%%
\be
\tilde\sigma\longrightarrow \tilde\sigma -2b\, \chi
\ee
%%%%%
under the dual residual gauge transformations.  However, if there
is no magnetic charge, and thus $[\chi]_{\theta=0}^{\theta=\pi}=0$,
then the angular momentum $\widetilde J$ calculated using (\ref{tangmom})
will be invariant also under (\ref{dualtrans}).

  It should be noted also that if we are able to make a gauge transformation
of the form (\ref{chiAtrans}) that sets $\chi$ to zero on the $z$ axis,
then the gauge-invariant expression (\ref{tangmom}) for the angular
momentum of an electrically-charged solution will coincide with the
expression, in general gauge-dependent, following from (\ref{angmom}).

\section{Mass of the Kerr-Newman-Melvin Black Holes}

   In this section, we shall be describing an approach to calculating the
mass of the magnetised black holes by means of a dimensional reduction to
three dimensions.  In order to avoid a profusion of annotations on
the three-dimensional equations we shall, {\it in this section only},
adopt the convention that four-dimensional quantities are denoted with
hats, while three-dimensional ones are unadorned.

\subsection{Hamiltonian formalism}

The original four-dimensional theory is given by
%%%%%%
\begin{equation}
\hat I =\frac1{16\pi G_4}\int_{\hat{M}}
\Big(\hat{R} - \hat{F}^2\Big)\,\sqrt{-\hat g}\, d^4 x
+\frac1{8\pi G_4}
\oint_{\partial \hat{M}}\hat{K}\, \sqrt{|\hat \gamma|}\, d^3x\,,
\end{equation}
%%%%%
where $\hat K$ is the extrinsic curvature of the three-dimensional 
boundary $\del\hat M$, which
has the induced metric $\hat\gamma_{\mu\nu}$. Upon dimensional
reduction on a circle using the standard Kaluza-Klein ansatz
%%%%%
\be
d\hat s_4^2 = e^{2\varphi}\, ds_3^2 + e^{-2\varphi}\, (d\phi + 2\cA)^2
\,,\qquad \hat A= A + \chi\, (d\phi+2 \cA)\,,\label{kkans2}
\ee
%%%%%
we obtain the three dimensional theory
%%%%%
\bea
I&=&\frac1{16\pi G_3}\int_{M}
\Big(R-2\Box\varphi-2(\partial\varphi)^2-2e^{2\varphi}(\partial\chi)^2-
e^{-4\varphi}{\cal F}^2-e^{-2\varphi}F^2\Big)\, \sqrt{-g}\, d^3 x\nn\\
&&+\frac1{8\pi G_3}\oint_{\partial M}(K+n^{\mu}\partial_{\mu}\varphi)
\,\sqrt{|\gamma|}\, d^2x \,,
\qquad G_4=(\Delta\phi)\,  G_3\,,
\eea
%%%%%
where $\Delta\phi$ is the period of the reduction coordinate $\phi$.
In the following, we set $G_4=1$, and therefore $G_3=1/(\Delta\phi)$.
After integration by parts,
%%%%%
\be
I=\frac{\Delta\phi}{16\pi}\int_{M}\Big(R-2(\partial\varphi)^2-
2e^{2\varphi}(\partial\chi)^2-e^{-4\varphi}{\cal F}^2-
e^{-2\varphi}F^2\Big)\, \sqrt{-g}\, d^3x
+\frac{\Delta\phi}{8\pi}\oint_{\partial M}K\, \sqrt{|\gamma|}\, d^2x\,.
\ee
%%%%%
Adding Lagrange multipliers $4d\psi\wedge(F-2\chi{\cal F})+
 4d\sigma\wedge {\cal F}$ and eliminating $F$ and ${\cal F}$, we arrive at
the dualised
Lagrangian describing three-dimensional gravity coupled to a sigma model
%%%%%
\be
I=\frac{\Delta\phi}{16\pi}\int_{M}\Big(R-2\Sigma_{AB}(\phi)\, 
  \partial\phi^A\partial\phi^B\Big)\, \sqrt{-g}\, d^3x
 +\frac{\Delta\phi}{8\pi}\oint_{\partial M}K\, \sqrt{|\gamma|}\, d^2x\,,
\ee
%%%%%
where $\phi^A$ represents all the scalars. The sigma-model metric
is
%%%%%
\be
d\Sigma^2=d\varphi^2+e^{2\varphi}(d\chi^2+d\psi^2)+
e^{4\varphi}(d\sigma-2\chi d\psi)^2\,.
\ee
%%%%%

      As stated before, the dualised action differs from the original one
by a total derivative term, and this will modify the definition of
energy\footnote{It is easy to see this in the Wald procedure, where
adding a total derivative term $d\nu$ to the Lagrangian will shift the
canonical charge associated with Killing vector $\xi$ by $i_{\xi}\nu$.}.
The original Lagrangian cannot easily be used to calculate the energy
because the corresponding Hamiltonian contains terms
such as $\oint_{S_{\infty}} A d\psi$ and 
$\oint_{S_{\infty}} {\cal A} d\sigma$,
whose evaluation is unclear.  However, these term are absent in the
dualised Lagrangian, rendering the calculation more well defined.  
We shall therefore carry out our calculations, 
and give a thermodynamic interpretation, using the 
dualised form of the Lagrangian.  This can be viewed as a choice of
regularisation scheme for giving a definition of mass that is 
applicable in the rather unusual asymptotic geometry of the magnetised
black hole solution.

   In the ADM decomposition, the three dimensional metric is recast into
the form
%%%%%
\be
ds^2_3=-N^2dt^2+h_{ij}(dx^i+N^idt)(dx^j+N^jdt)\,.
\ee
%%%%%
It follows that the Hamiltonian defined on the constant $t$ surface takes
the form \cite{Brown:1990fk}
%%%%
\be
H=\int_{\Sigma^t}d^2x(N{\cal H}+N^i{\cal H}_i)-
\oint_{S_{\infty}^t}dx\sqrt{\sigma}[\frac{\Delta\Phi}{8\pi }N k+
\frac{2}{\sqrt{h}}N^iP_{ij}n^j]\,,
\ee
%%%%%
where ${\cal H}$ and ${\cal H}_i$ are the total Hamiltonian constraint
and the momentum constraint. Using the extrinsic curvature $K_{ij}$ of 
$\Sigma^t$, the momentum $P^{ij}$, conjugate to $h_{ij}$, can be expressed as
%%%%%
\be
P_{ij}=\frac{\Delta \phi}{16\pi}\sqrt{h}(K_{ij}-Kh_{ij}),\qquad 
K_{ij}=\frac{1}{2N}(\dot{h}_{ij}-2\nabla_{(i}N_{j)})\,,
\ee
%%%%%
where $\nabla_i$ is defined with respect to $h_{ij}$. $S_{\infty}^t$, 
defined at $t={\rm const}$ and $r=\infty$, is a hypersurface 
inside $\Sigma^t$ with outward unit normal vector $n^i$. The
quantity 
$k \equiv h^{ij}\nabla_i n_j$ is the trace of the extrinsic 
curvature of $S_{\infty}^t$.
In general, the above expression for the Hamiltonian diverges. To 
obtain a
meaningful result, we must regularize the Hamiltonian by making
a subtraction in the surface term:
%%%%%
\be
H=\int_{\Sigma^t}d^2x(N{\cal H}+N^i{\cal H}_i)-\oint_{S_{\infty}^t}
dx\sqrt{\sigma}[\frac{\Delta\phi}{8\pi }N(k-k_0)+
\frac{2}{\sqrt{h}}N^ip_{ij}n^j]\,,
\label{Ham3}
\ee
%%%%%
where $k_0$ is the extrinsic curvature of $S_{\infty}^t$ embedded in
a certain two-dimensional reference background.

\subsection{Mass of the Kerr-Newman black hole}

Before computing the mass of the Kerr-Newman-Melvin black hole, we first
show how the three-dimensional Hamiltonian we have derived reproduces 
the standard 
mass for the Kerr-Newman black hole.

On shell, we have ${\cal H}={\cal H}^i=0$, and the Hamiltonian receives
contributions only from the boundary terms.
According to the reduction ansatz (\ref{kkans2}), the three-dimensional
metric induced from the four-dimensional Kerr-Newman black hole is given by
%%%%%
\bea
&&ds^2_{3{\rm KN}} =-
    \Delta\sin^2\theta dt^2+\Sigma\, \sin^2\theta
\, (\frac{dr^2}{\Delta}+ d\theta^2)\,,\label{gkn}\\
&&\rho^2=r^2+a^2 \cos^2\theta\,,\quad \Delta=r^2-2mr + a^2 + q^2\,,\quad
\Sigma=(r^2+a^2)^2 - a^2\Delta \sin^2\theta\,.\nn
\eea
%%%%%
This 3-metric is static, and so $p^{ij}=0$. The extrinsic curvature of
$S^t_{r=r_0}$ in $\Sigma^t$ can be computed, giving
%%%%%
\be
k = \fft1{\sqrt{\sigma}}\, \fft{\del\sqrt{\sigma}}{\del n}
=
\frac1{2\sin\theta}\sqrt{\frac{\Delta}{\Sigma}}
\frac{\partial_r\Sigma}{\Sigma}\Big|_{r=r_0}\,,
\ee
%%%%%
where $\sigma=g_{\theta\theta}$ is the determinant of the 1-dimensional
boundary metric and $\del/\del n$ is the derivative with respect to the
unit normal in the radial direction at $r=r_0$.

To compute $k_0$, we recall that the reference metric for the
four-dimensional Kerr-Newman black hole is the four-dimensional Minkowski
metric, which upon dimensional reduction gives rise to the three-dimensional
reference metric
%%%%%
\be
ds^2=-
 R^2\sin^2\theta dt^2+ R^4\, \sin^2\theta\,
\Big(\frac{dR^2}{R^2}+ d\theta^2\Big)\,.\label{refmet}
\ee
The calculation of $k_0$ requires us to embed $S^t_{r=r_0}$ into the above
background in such a way that the metric on $S^t_{r=r_0}$ induced from
the reference metric should be isometric to the metric on $S^t_{r=r_0}$
induced from $\Sigma^t$.  Thus the $t=\,$constant boundary at $R=R_0$
in the reference metric should be matched to the $t=\,$constant boundary
at $r=r_0$ in the reduction of the Kerr-Newman metric, implying
%%%%%
\be
R_0^4 = \Sigma\Big|_{r=r_0}\,.
\ee
%%%%%
This gives
%%%%%
\be
k_0 = \fft1{\sqrt{\sigma_0}}\, \fft{\del\sqrt{\sigma_0}}{\del n}
= \fft{2}{\sqrt{\Sigma}\, \sin\theta}\Big|_{r=r_0}\,,
\ee
%%%%%
where $\sigma_0= g_{\theta\theta}$ is the determinant of the 1-dimensional
boundary metric in the reference metric (\ref{refmet}).
Bearing in mind that the azimuthal coordinate $\phi$ has period
$\Delta\phi=2\pi$ in the Kerr-Newman metric, we therefore find 
from (\ref{Ham3}) that
%%%%%
\be
E_{\rm KN}=-\frac1{4}\oint_{S_{\infty}^t}dx\sqrt{\sigma}N(k-k_0)=m\,,
\label{knmass}
\ee
%%%%%
which reproduces the mass for the Kerr-Newman black hole.

\subsection{Mass of the Kerr-Newman-Melvin black hole}

We now turn to the calculation of the the mass of the Kerr-Newman-Melvin
black hole. The calculation closely resembles the previous case,
since the dimensionally-reduced 3-metric of the Kerr-Newman-Melvin black hole
is identical to that for the Kerr-Newman case, given in (\ref{gkn}).  The
three-dimensional evaluation of the mass differs in only one respect,
namely that the period $\Delta\phi$ of the azimuthal coordinate is
no longer $2\pi$, and so the mass is now given by
%%%%%
\be
E_{\rm KNM}=\fft{\Delta\phi}{2\pi}\,  m\,.\label{knmmass}
\ee
%%%%%

\subsection{Euclidean action}

The Lorentzian action is related to the Hamiltonian by
%%%%%
\be
I=\int d^4x\Big(P^{ij}\dot{h}_{ij}+\pi_A\dot{\phi}^A\Big)-\int dt H\,.
\ee
%%%%%
Therefore, for stationary solutions, the Euclidean action is given by
%%%%%
\be
I_E=\beta H\,.
\ee
%%%%%
The Hamiltonian will include a
contribution from the horizon. The total on-shell Hamiltonian
then takes the form \cite{Brown:1990fk}
%%%%%
\be
H=-\oint_{S_{\infty}^t}dx\sqrt{\sigma}[\frac{\Delta\phi}{8\pi }
N(k-k_0)+\frac{2}{\sqrt{h}}N^ip_{ij}n^j]-
\oint_{S^t_H}dx\sqrt{\sigma}[\frac{\Delta\phi}{8\pi }n^i\partial_iN-
\frac{2}{\sqrt{h}}N^ip_{ij}n^j]\,.
\ee
%%%%%
The first is equal to the energy $E$, and the second term gives rise to
$-TS$. Thus the Euclidean action of the dualised action is equal to
the Helmholtz free energy $F=E-TS$, suggesting that the dualised action
provides a canonical ensemble description for black-hole thermodynamics.

\section{Conserved Charges for Kerr-Newman-Melvin Black Holes}

\subsection{Kerr-Newman black holes}

   Before turning to the magnetised black hole metric, let us first
illustrate the three-dimensional procedure for calculating the conserved
charges by considering the original four-dimensional Kerr-Newman solution
itself, for which the reduction to three dimensions 
gives\footnote{Expressions for the three-dimensional fields $\psi$, 
$\chi$ and $\sigma$ can be found from those in \cite{gimupo}, by specialising
to the case where the external magnetic field is set to zero.  Note that
some sign conventions in \cite{gimupo}, associated with the 
definition of the orientation of the 2-spheres, differ from ours.}

%%%%%
\bea
d\bar s^2_3 &=&\sin^2\theta \Sigma(\frac{dr^2}{\Delta}+ d\theta^2)-
    \Delta\sin^2\theta dt^2\,,\nn\\
\chi &=& \fft{a q r \sin^2\theta}{\rho^2}\,,
\qquad \psi= -\fft{q(r^2+a^2)\cos\theta}{\rho^2}
\,,\nn\\
\sigma &=& -\fft{2 a m \cos\theta[r^2(3-\cos^2\theta) + a^2(1+\cos^2\theta)]}{
\rho^2} - \fft{2a^3 q^2 r \cos\theta \sin^4\theta}{\rho^4}\,,
\label{kn3dim}
\eea
%%%%
where
%%%%%
\be
\rho^2=r^2+a^2 \cos^2\theta\,,\qquad \Delta=r^2-2mr + a^2 + q^2\,,\qquad
\Sigma=(r^2+a^2)^2 - a^2\Delta \sin^2\theta\,.
\ee
%%%%%
For the mass, we saw that (\ref{knmass}) gave the expected result
%%%%%
\be
E=m\,.
\ee
%%%%%
The angular velocity and the electrostatic potential on
the horizon are given, in the three-dimensional calculation, by
%%%%%
\be
\Omega= -2 i_k\bar {\cal A}\Big|_{r=r_+} =\fft{a}{r_+^2+a^2}\,,\qquad
  \Phi_H= - i_k \bar A\Big|_{r=r_+} = \fft{q r_+}{r_+^2 + a^2}\,.
\ee
%%%%%
The electric
charge and angular momentum, given by (\ref{charge}) and (\ref{angmom}), 
are
%%%%%
\be
Q=\ft12
 \Big[\psi\Big]_{\theta=0}^{\theta=\pi}\ = q\,,\qquad
J= \ft14
 \Big[\sigma\Big]_{\theta=0}^{\theta=\pi} \  = am\,,
\qquad
\widetilde J= \ft14
 \Big[\tilde\sigma\Big]_{\theta=0}^{\theta=\pi} \  = am\,,
\label{knQJ}
\ee
%%%%%
since in this Kerr-Newman example the period $\Delta\phi$ of the azimuthal
coordinate $\phi$ is $2\pi$.  

   As we discussed in section 3.2, the expression for $J$ 
given in (\ref{angmom}) is not invariant under the residual gauge
transformations
%%%%%
\be
\chi\longrightarrow \chi+c\,,\qquad \sigma\longrightarrow 
\sigma + 2 c\, \psi\,,\qquad \psi\longrightarrow \psi\,,
\label{ctrans}
\ee
%%%%%
and in fact, from (\ref{Jtrans1}), we will have
%%%%%
\be
J\longrightarrow J + c\, Q
\ee
%%%%
in this case.  Thus the fact that $J$ in (\ref{knQJ}) has turned out to
give the correct result for the angular momentum is 
a consequence of happy choice of gauge; as can be seen from (\ref{kn3dim}), 
it is the one in which 
$\chi$ goes to zero on the axis at $\theta=0$ and $\theta=\pi$, 
thus implying that it is regular there.  Furthermore, it goes to zero
at infinity.  

  By contrast, as we discussed in section 3.2, the expression (\ref{tangmom})
for $\widetilde J$ {\it is} invariant under the residual gauge transformations
(\ref{ctrans}), and so it is not subject to the same ambiguities.

   Finally, we remark that a simple way to calculate the angular 
momentum of a dyonically charged Kerr-Newman black hole is to perform
first a duality transformation to the case with purely electric charge,
then and use the gauge-invariant expression (\ref{tangmom}). 

\subsection{Charges and thermodynamics for the Kerr-Newman-Melvin metrics}

  Here, we use the Kaluza-Klein formalism of section 3 to
calculate the electric charge and angular momentum for the Kerr-Newman-Melvin
metrics.  These are again given by (\ref{charge}) and (\ref{angmom}) or
(\ref{tangmom}), where expressions for
the scalar fields $\sigma$, $\chi$ and $\psi$ can be found 
in appendices A and B of \cite{gimupo}.  The period of $\phi$, determined
by the requirement of there being no conical singularity on the axis
at $\theta=0$ and $\theta=\pi$, is now given by \cite{gimupo}
%%%%%
\be
\Delta\phi= 2\pi \Big[1+\ft32 q^2 B^2 + 2aqm B^3 +
                     (a^2 m^2 +\ft1{16} q^4)B^4\Big]\,.\label{Deltaphi}
\ee
%%%%

   It is straightforward to see
that the expression for $\chi$ given in \cite{gimupo} is non-vanishing on
the axis at $\theta=0$ and $\theta=\pi$: we have
%%%%%
\be
\chi|_{\theta=0} = \chi|_{\theta=\pi} = \gamma\equiv \fft{\pi B}{4\Delta\phi}\,
[12 q^2 + 24 a m q B + (q^4 + 16 a^2 m^2) B^2]\,.
\ee
%%%%%%
It is therefore natural, in the light of the previous calculation for the 
Kerr-Newman black hole, to make a gauge transformation of the form
(\ref{ctrans}) with $c=-\gamma$ before evaluating the gauge-dependent 
expression (\ref{angmom}) for the angular momentum.  Assuming that 
we do this, we then find
%%%%%
\bea
Q &=& q + 2am B - \ft14 q^3 B^2\,,\label{QJ}\\
J&=& \widetilde J= 
a m - q^3 B - \ft32 am q^2 B^2 -\ft14 q B^3 (8 a^2 m^2 + q^4)
-\ft1{16} am B^4 (16 a^2 m^2 + 3 q^4) \,, \nn
\eea
%%%%%
(The calculation of $Q$ is discussed in \cite{gimupo}.)  By having chosen
the gauge where $\chi$ vanishes on the axis, we obtain the same expression
$J$ for the angular momentum as we get from the gauge-invariant 
expression $\widetilde J$ given by (\ref{tangmom}). 

  For the mass, we see from (\ref{knmmass}) that result for the
Kerr-Newman-Melvin metric
will be just the usual factor $m$, however now scaled by a factor of
$(\Delta\phi)/(2\pi)$, where $\Delta\phi$ is the period of the
azimuthal angle $\phi$, given in (\ref{Deltaphi}).
Thus we find
that the mass is given by
%%%%%
\be
E=m \Big[1+\ft32 q^2 B^2 + 2aqm B^3 +
                     (a^2 m^2 +\ft1{16} q^4)B^4\Big]
     \,.\label{kmmass}
\ee
%%%%%

  The area $A_H$ of the outer horizon
 and the surface gravity $\kappa$ can be straightforwardly calculated
from the Kerr-Newman-Melvin metrics given in \cite{gimupo}, leading to
%%%%%
\bea
A_H&=&4 \pi \Big(1+a^2 m^2 B^4+2 a mB^3 q+\frac{3 B^2q^2}{2}+
\frac{B^4 q^4}{16}\Big) \Big(a^2+(m+\sqrt{m^2-a^2-q^2})^2\Big)\,,\nn\\
\frac{\kappa}{8\pi}&=&\frac{\sqrt{m^2-a^2-q^2}}{8\pi
       \Big(a^2+(m+\sqrt{m^2-a^2 -q^2})^2\Big)}\,.\label{area}
\eea
%%%%%

  Assuming for now that we hold the external magnetic field $B$ fixed, we
can expect that the first law should take the form
%%%%%
\be
dE=\frac{\kappa}{8\pi}dA_H+\Omega dJ+\Phi dQ\,,\label{firstlaw}
\ee
%%%%%
where $\Omega=\Omega_H-\Omega_\infty$ is the difference between the
angular velocity of the horizon
and the angular velocity at infinity, and $\Phi=\Phi_H-\Phi_\infty$ is
the potential difference
between the horizon and infinity.  Because of subtleties associated with the
asymptotic structure of the Kerr-Newman-Melvin metrics at infinity
it is not obvious
how to calculate $\Omega_\infty$ and $\Phi_\infty$.   We can, however,
proceed by using our results
above for the other thermodynamic quantities, and then
seeking solutions for $\Phi$ and $\Omega$ such that the first law
(\ref{firstlaw}) holds.  We find that solutions do indeed exist.  This is
in fact non-trivial, since with three independent parameters being varied
in (\ref{firstlaw}) we have three equations for the two unknowns
$\Omega$ and $\Phi$.\footnote{The fact that a solution exists for $\Omega$
and $\Phi$ also provides non-trivial support for the validity
of our expression (\ref{kmmass}) for the mass of the Kerr-Newman-Melvin
solution.}  The solutions for $\Omega$ and $\Phi$ are rather
complicated, and we shall not present them in detail here.  Later, we
shall present leading-order terms in $\Omega$ and $\Phi$ in useful
approximations.

  Firstly, however, we remark that we can also allow $B$ to become an
additional thermodynamic variable in the first law, which will now be
generalised to
%%%%%
\be
dE=\frac{\kappa}{8\pi}dA_H+\Omega dJ+\Phi dQ - \mu dB\,,\label{genfirstlaw}
\ee
%%%%%
where $\mu$ has the interpretation of being the magnetic moment of the
system.  (Analogous expressions
have been obtained by for the case of Einstein-Dilaton-Maxwell
theory in the Kaluza-Klein case by  Yazadjiev \cite{yaz}.) 
Again, it is non-trivial that a solution for $\mu$ exists.  Having
obtained $\mu$, which is also rather complicated in general, it is
straightforward to verify that the various thermodynamic quantities
satisfy the Smarr-like relation
%%%%%
\be
E= \frac{\kappa}{4\pi} A_H + 2\Omega J  + \Phi Q + \mu B\,.
\ee
%%%%

 As we mentioned above, the solutions for $\Omega$, $\Phi$ and $\mu$
are rather complicated in general.  It is instructive to look at
the leading-order
forms of these quantities.  Up to linear order in $q$, we find
%%%%%
\be
\mu = a q(1+a^2m^2 B^4) + {\cal O}(q^2)\,.\label{muexp}
\ee
%%%%%
To linear order in $B$, we find
%%%%%
\bea
\Omega &=& \fft{a}{r_+^2 +a^2} - \fft{2 q B r_+}{r_+^2 +a^2} +
       {\cal O}(B^2)\,,\nn\\
\Phi &=& \fft{q r_+}{r_+^2+a^2} +
  \fft{3 a q^2 B}{(r_+^2+a^2)} + {\cal O}(B^2)
\,.\label{OmPhi}
\eea
%%%%%
Note that from (\ref{muexp}) we have, to lowest order, that 
$\mu = J q/m$, reproducing the gyromagnetic ratio $g=2$ as found 
by Carter \cite{Carter}.
We also see that  the second term in the expression for $\Omega$ in(
\ref{OmPhi}) agrees with the standard formula for the Larmor precession 
frequency  $\Omega_L = \mu B/J$, in the limit that we may approximate 
$r_+$ by $2m$.

\subsection{The case $q=-a m B$}

It was shown in \cite{gimupo} that in general the magnetised Kerr-Newman
black holes have an ergoregion that extends out to infinity close to the
axis of rotation.  A special case arises if the charge parameter $q$
of the original Kerr-Newman solution is chosen to satisfy \cite{gimupo}
%%%%%
\be
q=-a m B\,,\label{qspec}
\ee
%%%%%
where $m$ and $a$ are the mass and rotation parameters of the 
Kerr-Newman metric.  Under these circumstances we find that the 
conserved charge and angular momentum, given in general by (\ref{QJ}),
simplify considerably, and become
%%%%%
\bea
Q &=& a m B\, \sqrt{\fft{\Delta\phi}{2\pi}}= 
     - q\,\sqrt{\fft{\Delta\phi}{2\pi}}= 
- Q_0\, \sqrt{\fft{\Delta\phi}{2\pi}}\,,\nn\\
J &=& a m\, \fft{\Delta\phi}{2\pi} = J_0\, \fft{\Delta\phi}{2\pi}\,,
\eea
%%%%%
where $Q_0=q$ and $J_0=am$ are the conserved charge and angular momentum
of the original Kerr-Newman solution.
The period of the azimuthal coordinate is now
%%%%%
\be
\fft{\Delta\phi}{2\pi}= (1 + \ft14 a^2 m^2 B^2)^2\,.
\ee
%%%%%
The area of the event horizon, given in general by (\ref{area}), 
can now be written as
%%%%%
\be
A_H = A^0_H\, \fft{\Delta\phi}{2\pi}\,,
\ee
%%%%%
where $A^0_H$ is the area of the event horizon of the 
Kerr-Newman black hole.  Of course we still also have, from (\ref{knmmass}), 
that the mass is given by
%%%%%
\be
E= E_0\, \fft{\Delta\phi}{2\pi}\,,
\ee
%%%%%
where $E_0=m$ is the mass of the Kerr-Newman black hole, while the 
surface gravity $\kappa$ is, as always, just equal to its value in
the Kerr-Newman solution (see equation (\ref{area})). 

\section{Energy Minimisation}

  Defining $E_0=m$ and $J_0=a m$ as the mass and the angular momentum of
the Kerr-Newman black hole (i.e. the $B=0$ specialisation), we may eliminate
$q$ between the expressions for $Q$ and $E$ in (\ref{QJ}) and (\ref{kmmass}),
thereby obtaining an equation that determines $E$ in terms of $Q$, $J_0$
and $B$:
%%%%%
\bea
&&E^3 - E^2 (17+ 3 B^4 J_0^2) E_0 +
\ft12 E(160-192 B^4 J_0^2 + 6 B^8 J_0^4 + 136 B^3 J_0 Q - 11 B^2 Q^2)E_0^2\nn\\
&&-\Big(64 + 48 B^4 J_0^2 + 12 B^8 J_0^4 + B^{12} J_0^6 -128 J_0 Q -
  32 B^7 J_0^3 Q + 68 B^2 Q^2 + 17 B^6 J_0^2 Q^2 \nn\\
&&\qquad- 2 B^5 J_0 Q^3 +
\ft1{16} B^4 Q^4\Big) E_0^3=0\,.
\eea
%%%%%
Extremising $E$ with respect to $Q$, while holding $E_0$, $J_0$ and $B_0$
fixed then implies
%%%%%
\bea
E =\fft{(512 B J_0 + 128 B^5 J_0^3 - 544 Q - 136 B^4 J_0^2 Q +
  24B^3 J_0 Q^2 - B^2 Q^3) E_0}{4(11 Q- 68 B J_0)}\,.
\eea
%%%%%
From these equations we can now obtain expressions for $\bar E$ and
$\bar Q$, the values of $E$ and $Q$ at the extremum, as function of
$E_0$, $J_0$ and $B$.  For $Q$, we find that $\bar Q$ is given by the roots
of the factorised polynomial $P(Q)=P_1(Q) P_2^2(Q)$, where
%%%%%
\bea
P_1&=& B^2 Q^3 - 12 B^3 J_0 Q^2 + 16(4+B^4 J_0^2)(3Q-4B J_0)\,,\nn\\
P_2&=& B^2 Q^3 - 30 B^3 J_0 Q^2 -4(392-75 B^4 J_0^2) Q +
 8 B J_0(588-125 B^4 J_0^2)\,.
\eea
%%%%%
Expanding around $B=0$
we find just one real root for $P_1(Q)=0$, giving
%%%%%
\bea
\bar Q &=& \ft43 B J_0 + \ft{8}{81} B^5 J_0^3 + \cdots\,,\nn\\
\bar E &=& E_0 + \ft13 E_0 B^4 J_0^2 +\cdots\,.
\eea
%%%%%
For $P_2(Q)=0$ we find three real roots, with
%%%%%
\bea
\bar Q &=& 3 B J_0+\cdots\,,\nn\\
\bar E &=& 8 E_0 -\ft34 E_0 B^4 J_0^2 +\cdots
\eea
%%%%%
or
%%%%%
\bea
\bar Q &=& \pm \fft{28\sqrt2}{B} +\fft{27}{2}\, B J_0 \mp
 \fft{21}{32\sqrt 2}\, B^3 J_0^2 +\cdots\,,\nn\\
\bar E &=& -48 E_0 \mp 7\sqrt2 E_0 B^2 J_0 +\fft{15}{8} E_0 B^4 J_0^2
   +\cdots\,.
\eea
%%%%%

\section{Further Properties of Uncharged Black Holes}

   In this section, we explore various properties of
magnetised Kerr-Newman black holes in the special case where the physical
charge $Q$ vanishes.

\subsection{Angular momentum of uncharged magnetised black holes}

  Using the expressions (\ref{QJ}) for the physical charge $Q$ and the
angular momentum $J$ of a magnetised Kerr-Newman black hole, we may express
$J$ in terms of $J_0=am$ and $B$ in the case that $Q$ is required to be zero.
Note that here $J_0$ is the angular momentum of the unmagnetised
Kerr-Newman seed solution. We find that $J$, $J_0$ and $B$ are then related
by
\be
B^4 J^3 + B^4 J_0\, (79+ 3 B^4 J_0^2)\, J^2 -
(256 -944 B^4 J_0^2 + 248 B^8 J_0^4 - 3B^{12} J_0^6)\, J +
  J_0\, (4+B^4 J_0^2)^4=0\,.\label{xxexp}
\ee
%%%%%
Expanding in powers of $B$, the branch that reduces to $J=J_0$ in the case
that $B$ vanishes gives
%%%%%
\be
J= J_0 + 5 B^4\, J_0^3 + 21 B^8\, J_0^5 + 94 B^{12}\, J_0^7 +
  454 B^{16}\, J_0^9 +\cdots\,.
\ee
%%%%%
In order that $J$ remain real the product $B^2\, J_0$ should not
exceed a maximum value, given by
%%%%%
\be
B^2\, J_0\big|_{\rm max} = \fft{2}{3\sqrt3}\,.
\ee
%%%%%
This corresponds to
%%%%%
\be
J\big|_{\rm max} = \fft{128}{27}\, J_0\big|_{\rm max}
  = \fft{256}{81\sqrt3\, B^2}\,.
\ee
%%%%%

\subsection{Meissner effect for extremal black holes}

  The electromagnetic field in the magnetised Kerr-Newman solution takes the
form $A=\bar A +\chi\, (d\phi +2 \bar\cA dt)$, as in (\ref{kkred}),
where the various
quantities may be found in appendix B of \cite{gimupo}.  The magnetic flux
threading the upper hemisphere $S^2_+$ of the horizon is given by
%%%%%
\be
\cF_H = \fft1{4\pi}\, \int_{S^2_+} F = \fft{\Delta\phi}{4\pi}\,
  \Big[\chi\Big]_{\theta=\ft12\pi}^{\theta=\pi}\,.\label{flux}
\ee
%%%%%

Consider the case where the physical charge $Q$ on the black hole vanishes.
From (\ref{QJ}), this is achieved if the magnetic field is given by
$B=B_\pm$ where
%%%%%
\be
B_\pm = \fft{2}{q^3}\, \Big[2 a m \pm\sqrt{4 a^2 m^2 + q^4}\Big]\,.
\label{zerocharge}
\ee
%%%%%
Suppose furthermore that the black hole is extremal,
which means that the inner and outer horizons at $r=r_\pm$ coincide at
%%%%
\be
r_\pm= m\,,\qquad m=\sqrt{q^2+a^2}\,.\label{extremal}
\ee
%%%%%
Inserting the zero-charge condition (\ref{zerocharge})
and the extremality condition (\ref{extremal}) into the expression for
$\chi$ given in \cite{gimupo}, we find that $\chi$ is constant on the
horizon, and it is given by
%%%%%
\be
\chi\big|_H = \pm \fft{q^3}{2(q^2+2a^2)}\,.
\ee
%%%%%
Evidently therefore, from (\ref{flux}), it follows that the magnetic
flux threading the upper hemisphere of the horizon is zero.
This is consistent
with much previous work (see \cite{Bicak} for a review).

\section{Angular Momentum in STU Supergravity}\label{stusec}

   In this section, we extend our earlier discussion of the conserved
angular momentum in Einstein-Maxwell theory to the case of the
four-dimensional STU model, which is ${\cal N}=2$ supergravity coupled
to three vector multiplets.  For our purposes, it suffices to focus
just on the bosonic sector of the theory.  The bosonic Lagrangian,
in the notation of \cite{cclp}, is
%%%%%
\bea
{\cal L}_4 &=& R\, {*\oneone} - \ft12 {*d\varphi_i}\wedge d\varphi_i
   - \ft12 e^{2\varphi_i}\, {*d\chi_i}\wedge d\chi_i - \ft12 e^{-\varphi_1}\,
\Big( e^{\varphi_2-\varphi_3}\, {* F_{\2 1}}\wedge F_{\2 1}\nn\\
&& + e^{\varphi_2+\varphi_3}\, {*  F_{\2 2}}\wedge F_{\2 2}
   + e^{-\varphi_2 + \varphi_3}\, {* \cF_\2^1 }\wedge \cF_\2^1 +
     e^{-\varphi_2 -\varphi_3}\, {*\cF_\2^2}\wedge \cF_\2^2\Big)\nn\\
&& + \chi_1\, (F_{\2 1}\wedge \cF_\2^1 +
                  F_{\2 2}\wedge \cF_\2^2)\,,
\label{d4lag}
\eea
%%%%%
where the index $i$ labelling the dilatons $\varphi_i$ and axions $\chi_i$
ranges over $1\le i \le 3$.  The four field strengths can be written in
terms of potentials as
%%%%%
\bea
F_{\2 1} &=& d A_{\1 1} - \chi_2\, d\cA_\1^2\,,\nn\\
F_{\2 2} &=& d A_{\1 2} + \chi_2\, d\cA_\1^1 -
    \chi_3\, d A_{\1 1} +
      \chi_2\, \chi_3\, d\cA_\1^2\,,\nn\\
\cF_\2^1 &=& d\cA_\1^1 + \chi_3\, d \cA_\1^2\,,\nn\\
\cF_\2^2 &=& d\cA_\1^2\,.
\eea
%%%%%

\subsection{Derivation of the conserved angular momentum}

  The conserved charge associated with a diffeomorphism $\xi$ can be calculated
using the standard Wald procedure.  Thus, we first calculate
$\delta{\cal L}(\Phi) =
d\Theta + \hbox{e.o.m. terms}$, where all the fields $\Phi$
are varied using the Lie derivatives $\delta\Phi ={\mathfrak L}_\xi\, \Phi$.  
For example,
for the metric we have $\delta g_{\mu\nu}= \nabla_\mu\xi_\nu +
\nabla_\nu\xi_\mu$, and for gauge potentials
$\delta A_\mu =\xi^\nu \nabla_\nu A_\mu + A_\nu\nabla_\mu\xi^\nu$.
In the standard way, we then define
%%%%%
\be
{\cal J}= \Theta -i_\xi{\cal L}\,,
\ee
%%%%%
where $i_\xi$ denotes the contraction of the vector
$\xi$ with the Lagrangian 4-form ${\cal L}$, as
defined in footnote \ref{idef}.
It follows that $d{\cal J}=0$ and hence we can write
%%%%%
\be
{\cal J}= -d{\cal P}\,.
\ee
%%%%%
After considerable algebra, we find that for the Lagrangian (\ref{d4lag})
we shall have
%%%%%
\be
{\cal P} = {\cal P}_{\rm Ein} + {\cal P}_{\rm Kin} + {\cal P}_{\rm CS}\,,
\label{Qsum}
\ee
%%%%%
where
%%%%%
\bea
{\cal P}_{\rm Ein} &=& {*d\xi}\,,\nn\\
{\cal P}_{\rm Kin} &=&
  e^{-\varphi_1+\varphi_2-\varphi_3}\, {*F}_{\2 1} \,
  \xi^\mu(A_{\mu\, 1} -\chi_2\, \cA_\mu^2)\nn\\
&&
 + e^{-\varphi_1+\varphi_2+\varphi_3}\, {*F}_{\2 2} \,
 \xi^\mu(A_{\mu 2} + \chi_2\, \cA_\mu^1 -\chi_3\, A_{\mu 1} +
  \chi_2\, \chi_3\, \cA_\mu^2)\nn\\
&&
  +e^{-\varphi_1-\varphi_2+\varphi_3}\, {*\cF}_{\2}^1 \,
 \xi^\mu(\cA_\mu^1 + \chi_3\, \cA_\mu^2)
  +e^{-\varphi_1-\varphi_2-\varphi_3}\, {*\cF}_{\2}^2 \,
 \xi^\mu\, \cA_\mu^2
  \,,\nn\\
{\cal P}_{\rm CS} &=& -\chi_1\, [(\xi^\mu \,A_{\mu 1}) d\cA_\1^1 +
   (\xi^\mu\, \cA_\mu^1) dA_{\1 1} +
  (\xi^\mu \,A_{\mu 2}) d\cA_\1^2 +
   (\xi^\mu\, \cA_\mu^2) dA_{\1 2}]\,.
\eea
%%%%%
Here ${\cal P}_{\rm Ein}$ is the contribution from the Einstein-Hilbert
term in (\ref{d4lag}), ${\cal P}_{\rm Kin}$ is the contribution from the
four kinetic terms for the gauge field strengths, and ${\cal P}_{\rm CS}$ is
the contribution from the Chern-Simons terms.

   We now make the spacelike dimensional reduction
%%%%%
\be
ds_4^2 = e^{-\varphi_4}\, d\bar s_3^2 + e^{\varphi_4}\,
   (d\phi + \bar\cB_\1)^2
\,,\label{metans}
\ee
%%%%%
and
%%%%%
\bea
A_{\1 1} &=& \bar A_{\1 1} + \sigma_1\, (d\phi + \bar\cB_\1)\,,\qquad
A_{\1 2} = \bar A_{\1 2} + \sigma_2\, (d\phi + \bar\cB_\1)\,,\nn\\
\cA_\1^1 &=& \bar\cA_\1^1 + \sigma_3\, (d\phi+\bar\cB_\1)\,,\qquad
\cA_\1^2 = \bar\cA_\1^2 + \sigma_4\, (d\phi+\bar\cB_\1)\,,\label{kkAred}
\eea
%%%%%
otherwise following the notation of the timelike reduction described
in \cite{cclp}.  In particular, in three dimensions the four reduced
field strengths
and the Kaluza-Klein field strength $\bar \cG_\2=d\bar \cB_\1$
 are re-expressed in terms of
scalar fields, by means of dualisations \cite{cclp}:
%%%%%
\bea
- e^{-\varphi_1 + \varphi_2 - \varphi_3 + \varphi_4}\, {\bar * \bar F_{\2 1}}
&=& d\psi_1 + \chi_3\, d\psi_2 - \chi_1\, d\sigma_3 - \chi_1\,
\chi_3\, d\sigma_4\,,\nn\\
- e^{-\varphi_1 + \varphi_2 + \varphi_3 + \varphi_4}\, {\bar * \bar F_{\2 2}}
&=& d\psi_2 -\chi_1\, d\sigma_4\,,\nn\\
- e^{-\varphi_1 - \varphi_2 + \varphi_3 + \varphi_4}\, {\bar * \bar \cF_{\2}^1}
&=& d\psi_3 - \chi_2\, d\psi_2 - \chi_1\, d\sigma_1 + \chi_1\,
\chi_2\, d\sigma_4\,,\nn\\
- e^{-\varphi_1 - \varphi_2 - \varphi_3 + \varphi_4}\, {\bar * \bar\cF_{\2}^2}
&=& d\psi_4 + \chi_2\, d\psi_1 - \chi_3\, d\psi_3 -
\chi_1\,d\sigma_2 + \chi_2\, \chi_3\, d\psi_2 \nn\\
&&- \chi_1\, \chi_2\, d\sigma_3 +
 \chi_1\, \chi_3\, d\sigma_1 - \chi_1\, \chi_2\, \chi_3\, d\sigma_4\,,\nn\\
e^{2\varphi_4}\, {\bar *\bar \cG_\2} &=& d\chi_4 + \sigma_1\, d\psi_1 +
    \sigma_2\, d\psi_2 +\sigma_3\, d\psi_3 +   \sigma_4\, d\psi_4\,.
\label{dualfields}
\eea
%%%%%

   We then find after some algebra that with $\xi=\del/\del\phi$
we shall have
%%%%%
\bea
{\cal P}_{\rm Ein} &=& (d\chi_4 + \sigma_1\, d\psi_1 +
  \sigma_2\, d\psi_2 + \sigma_3\, d\psi_3 + \sigma_4\, d\psi_4)\wedge d\phi
  + \cdots\nn\\
{\cal P}_{\rm Kin} &=& [-\sigma_1\, d\psi_1 -
  \sigma_2\, d\psi_2 - \sigma_3\, d\psi_3 - \sigma_4\, d\psi_4+
  \chi_1\, d(\sigma_1\, \sigma_3 + \sigma_2\, \sigma_4)]\wedge d\phi + \cdots
\nn\,,\\
{\cal P}_{\rm CS} &=& -\chi_1\, d(\sigma_1\, \sigma_3 + \sigma_2\, \sigma_4)
\wedge d\phi + \cdots\,,
\eea
%%%%%
where the ellipses denote terms that have vanishing pullback to 
the 2-sphere.
Thus, from (\ref{Qsum}) we conclude that ${\cal P}= d\chi_4\wedge d\phi
 +\cdots$, and so the conserved charge associated
with the Killing vector $\xi=(\Delta\phi/(2\pi))\, \del/\del\phi$ is given by
%%%%%
\be
J= \fft{1}{16\pi} \int_{S^2} {\cal P} = \fft{(\Delta\phi)^2}{32\pi^2}\,
  \int d\chi_4= \fft{(\Delta\phi)^2}{32\pi^2}\,
\Big[\chi_4\Big]_{\theta=0}^{\theta=\pi}\,.\label{Jexp}
\ee
%%%%%
(The $(\Delta\phi/(2\pi))$ factor in the choice of the Killing
vector takes account of the fact that angular momentum should be defined
with respect to a canonically-normalised azimuthal angle having
period $2\pi$.)

   This result is the analogue of the expression we derived in (\ref{angmom})
for the angular momentum for the Einstein-Maxwell black holes.  It
also has the same feature as in that case, of not being invariant under
gauge transformations of the electromagnetic potentials.  Specifically,
we have four abelian $U(1)$ gauge symmetries in the STU model, under which
the potentials transform as
%%%%%
\be
A_\1^{[i]} \longrightarrow A_\1^{[i]}{'} =
A_\1^{[i]} + d\lambda_i\,,\label{gaugetransi}
\ee
%%%%%
where $A_\1^{[i]}$ for $i=1$, 2, 3 and 4 denotes
$(A_{\1 1}, A_{\1 2}, \cA_\1^1, \cA_\1^2)$ respectively.  The subset of
gauge transformations where
%%%%%
\be
\lambda_i=\bar\lambda_i + c_i\, \phi\,,
\ee
%%%%
with $\bar\lambda_i$ being independent of $\phi$ and $c_i$ being constants,
preserve the form of the Kaluza-Klein reductions (\ref{kkAred}).  For these
gauge transformations, the three-dimensional gauge potentials and the
$\sigma_i$ fields therefore transform as
%%%%%
\be
\bar A_\1^{[i]}{'} = \bar A_\1^{[i]} - c_i\, \bar\cB_\1 + d\bar\lambda_i\,,
\qquad \sigma_i'=\sigma_i + c_i\,.\label{sigitrans}
\ee
%%%%%
From these, it follows that the quantities $d\bar A_\1^{i]} + \sigma_i\,
 d\bar\cB_\1$ from which three-dimensional field strengths
$\bar F_\2^{[i]}$ are constructed, and hence the three-dimensional field
strengths themselves, are inert under the gauge transformations.  This in
turn implies that the scalar fields $\psi_i$ are inert,
%%%%%
\be
\psi_i'=\psi_i\,.\label{psiitrans}
\ee
%%%%%
Finally, since $\bar\cF_\2$ is inert, it follows from (\ref{sigitrans}) and
(\ref{psiitrans}) that $\chi_4$ transforms as
%%%%%
\be
\chi_4' = \chi_4 -  \sum_i c_i\, \psi_i\,.\label{chi4trans}
\ee
%%%%%
Thus we see that under the gauge transformations (\ref{gaugetransi}),
the angular momentum $J$ defined in (\ref{Jexp}) transforms as
%%%%%
\be
J' = J -\fft{\Delta\phi}{8\pi}\, \sum_i c_i \, Q_i\,,
\ee
%%%%%
where
%%%%%
\be
Q_i = \fft{\Delta\phi}{4\pi}\, \Big[\psi_i\Big]_{\theta=0}^{\theta=\pi}
\ee
%%%%%
are the electric charges carried by the four field strengths.

   As in the Einstein-Maxwell case discussed earlier, we can derive a
different expression for the angular momentum, which {\it is} gauge invariant,
by performing dualisations of all the four gauge fields.  This is
easily done in the three-dimensional description, where it amounts to
sending $\sigma_i$, $\psi_i$ and $\chi_4$ to tilded quantities, defined by
%%%%%
\be
\tilde\sigma_i=\psi_i\,,\qquad \tilde \psi_i=\sigma_i\,,\qquad
\widetilde\chi_4 = \chi_4 + \sum_i \sigma_i\, \psi_i\,.\label{tilded}
\ee
%%%%%
Repeating the calculation of the angular momentum for the dualised theory
will give
%%%%%
\be
\widetilde J= \fft{(\Delta\phi)^2}{32\pi^2}\,
\Big[\widetilde\chi_4\Big]_{\theta=0}^{\theta=\pi}\,.\label{JJexp}
\ee
%%%%%
Using our results above for the gauge transformations of $\sigma_i$,
$\psi_i$ and $\chi_4$, it is easily seen that $\widetilde\chi_4$, and hence
$\widetilde J$, is gauge invariant.  We can then argue, in a manner analogous
to our argument in the Einstein-Maxwell case, that (\ref{JJexp}) would be
the appropriate expression to use if all four of the charges carried by the
gauge fields were electric.

\subsection{Angular momentum for the magnetised static STU-model black holes}

   In the case of the four-charge black holes in the STU model discussed
in \cite{cclp}, and with external fields in \cite{cvgiposa},
the field strengths numbered 1 and 3 carry
magnetic charges, whilst those numbered 2 and 4 carry electric charges.
It follows, therefore, that rather than using (\ref{JJexp}) directly
in order to calculate the angular momentum, we should first ``undualise''
the contributions associated with fields 1 and 3, meaning that
$\widetilde\chi_4$ in (\ref{JJexp}), which was defined in (\ref{tilded}),
 should be replaced by
$\widetilde\chi_4 - \sigma_1\, \psi_1-\sigma_3\, \psi_3$.
Thus the proposal for the angular momentum in this case is now
%%%%%
\be
J_e= \fft{(\Delta\phi)^2}{32\pi^2}\,
\Big[\chi_4 + \sigma_2\, \psi_2 +
       \sigma_4\, \psi_4\Big]_{\theta=0}^{\theta=\pi}\,.\label{JJ4che}
\ee
%%%%%
Substituting the expressions for the scalar fields obtained in \cite{cvgiposa},
we obtain the result
%%%%%
\be
J_e =\ft14 \Pi_q\, \sum_{i=1}^4 \fft{B_i}{q_i} + 
\ft1{16} \Pi_B\, \Pi_q\, \sum_{i=1}^4 \fft{q_i}{B_i}\,,
\label{JJtrue}
\ee
%%%%%
for the angular momentum of the magnetised 4-charge 
black holes
immersed in the background of the four external fields $B_i$, where $\Pi_B=
\prod_i B_i$ and $\Pi_q=\prod_i q_i$.   Here $q_i$ are the four charges of
the original static black hole solutions, prior to the magnetisation.
This expression reduces, as it should,
to $\widetilde J$ given in (\ref{QJ})
if we set all four charge parameters $q_i$ equal and set $a=0$ in (\ref{QJ}).

   An alternative way to calculate the angular momentum is to use the
four-charge analogue of the expression (\ref{angmom}) that we considered
in the Einstein-Maxwell case.  In the present context, this
amounts to starting from the expression (\ref{Jexp}), and then dualising
the contributions from fields 1 and 3 to take account of the fact that
they actually carry magnetic charges.  This gives
%%%%%
\be
J_m= \fft{(\Delta\phi)^2}{32\pi^2}\,
\Big[\chi_4 + \sigma_1\, \psi_1 +
       \sigma_3\, \psi_3\Big]_{\theta=0}^{\theta=\pi}\,.\label{JJ4chm}
\ee
%%%%%
Since (\ref{Jexp}) is gauge
dependent, it is then necessary to perform gauge transformations to
ensure that the four functions $(\sigma_1,\psi_2,\sigma_3,\psi_4)$ 
vanish on the axis at $\theta=0$ and $\theta=\pi$.  After doing this, we
obtain a result that agrees with the gauge-invariant one given in
(\ref{JJtrue}).

\section{Conclusions}

  In this paper we have obtained expressions for the energy and angular 
momentum  of magnetised Kerr-Newman black holes. We showed how
these quantities can be conveniently calculated by making a Kaluza-Klein
reduction of the four-dimensional Einstein-Maxwell theory, and the black
hole solutions, on the azimuthal coordinate $\phi$.  Using these 
expressions, we have 
verified the first law of thermodynamics  and  the associated Smarr 
formulae   for rotating black holes immersed in an external  magnetic
field.
We also extended the the first law to include variations of the magnetic 
field, and  hence we obtained the induced  magnetic moment.
In an attempt to make contact with some early work of Wald
in which the magnetic field was  treated at the test level, ignoring 
back-reaction, we have calculated the electric charge that minimises  
the energy,  holding the initial energy and angular momentum fixed and 
at constant  magnetic field. Our results resemble qualitatively those of 
Wald but differ quantitatively. Finally we extended our calculation of
the angular momentum to the case of the STU model of four-dimensional
${\cal N}=2$ 
supergravity coupled to three vector multiplets, in preparation for a 
future paper \cite{cvgiposa} on that subject.

\section*{Acknowledgements}

We are grateful to Mirjam Cveti\v c for useful conversations. 
The research of C.N.P. is supported in part by
DOE grant DE-FG03-95ER40917.

\end{document}